\documentclass[aps,prl,twocolumn,a4paper,10pt,notitlepage,footinbib,superscriptaddress,showpacs]{revtex4-1}%
\usepackage[english]{babel}
\usepackage[T1]{fontenc}
\usepackage{endnotes}
\usepackage{amssymb,amsmath,amsfonts}
\usepackage{textcomp}
\usepackage{graphicx,color}
\usepackage[utf8x]{inputenc}
\usepackage{float}
\usepackage{placeins}
\usepackage{multirow}
\usepackage{colortbl}
\usepackage{tabulary}
\usepackage{etoolbox}

\begin{document}

\title{Cascade or not cascade?
Energy Transfer and Elastic Effects in Active Nematics}

\author{Livio Nicola Carenza}
\affiliation{Dipartimento  di  Fisica,  Universit\`a  degli  Studi  di  Bari  and  INFN,  via  Amendola  173,  Bari,  I-70126,  Italy}

\author{Luca Biferale}
\affiliation{Dipartimento di Fisica and INFN, Universit\`a di Roma “Tor Vergata”, Via Ricerca Scientifica 1, 00133 Roma, Italy}

\author{Giuseppe Gonnella} 
\affiliation{Dipartimento  di  Fisica,  Universit\`a  degli  Studi  di  Bari  and  INFN,  via  Amendola  173,  Bari,  I-70126,  Italy}

\begin{abstract}
We numerically study the multi-scale properties of a $2d$ active gel to address the energy transfer mechanism. We find that activity is able to excite long-ranged distortions of the nematic pattern giving rise to spontaneous laminar flows and to a chaotic regime by further increasing the rate of active energy injection. By means of a scale-to-scale spectral analysis we find that the gel is basically driven by the local balancing between active injection and viscous dissipation, without any signal of non-linear hydrodynamical transfer and turbulent cascades. Furthermore, elasticity may qualitatively play an important role by transferring energy from small to larger scales through  nemato-hydrodynamic interactions.
\end{abstract}

\maketitle

\section{Introduction}

\begin{figure*}[bt]
\centering
\includegraphics[width=1.0\textwidth]{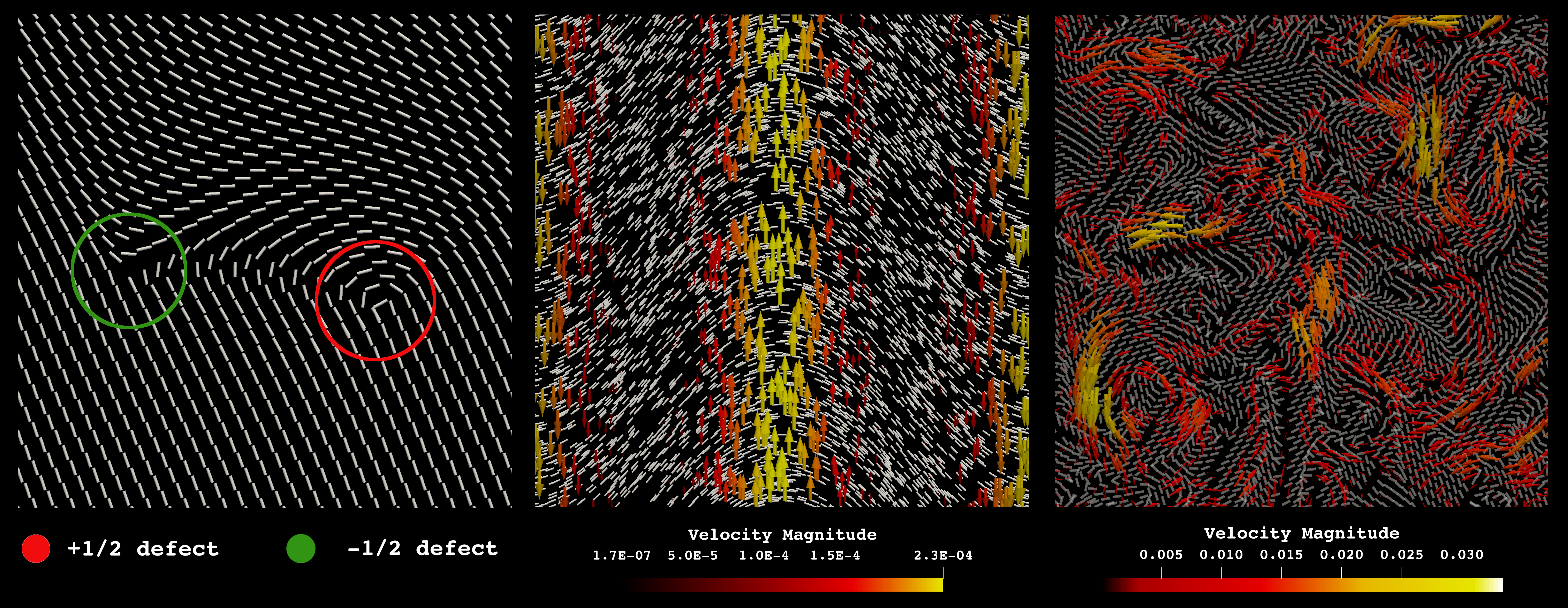}
\caption{\textbf{Dynamical regimes of active nematics.} Numerical simulations of an active nematics with elastic constant $K=0.04$ at varying the intensity of the activity $\zeta$. The picture is a summary of the three dynamical regimes occurring in active gels (whose spectral properties we discuss in this paper) as previously reported in~\cite{giomi2014_2,thampi2014_1,giomi2011,Urzay2017}. Left panel shows the configuration of the nematic field (white rods) during the relaxing dynamics in proximity of two oppositely charged semi-integer defects in the quiescent regime (\emph{i.e.} at low activity $\zeta=10^{-5}$). Central panel shows a configuration at $\zeta=5\times 10^{-4}$ in the spontaneous flow regime. The nematic field undergoes an activity-induced banding instability that breaks the rotational symmetry, producing bands of laminar flow (colored arrows) which is stronger where elastic deformations are more pronounced. Right panel shows a typical chaotic configuration at strong activity $\zeta= 5 \times 10^{-2}$, where rotational symmetry is statistically restored. This state is characterized by the formation of \emph{walls}, narrow regions of strong banding deformations, which eventually give rise to the enucleation of topological defects~\cite{bonelli2019}. Simulations are performed on a square grid of size $L=512$ and different portions of the systems are shown in the three panels.}
\label{fig1}
\end{figure*}

Active fluids are driven to an out-of-equilibrium state by the injection of energy at microscopic length-scales~\cite{ramaswamy2010,marc2013}. 
Systems of biological origin, such as cytoskeletal or bacterial suspensions, can be made \emph{active} 
by dispersing ATP or tuning oxygen concentration~\cite{wensink2012,guillamat2018}, while synthetic realizations (Janus particles~\cite{ebbens2014} or polyacrylic acid hydrogels~\cite{korevaar2020}) are able to convert chemical energy into motion. 
When activated, the individual constituents tend to arrange in a liquid-crystalline fashion, developing orientational (polar or nematic) order~\cite{sanchez2012}, in accordance to their intrinsic (vector or head-tail) symmetry.
The visco-elastic response of active gels to internal energy input has been widely analyzed in the past decade~\cite{kruse2004,marc2013}, as we briefly summarize in the following.
Varying the strength of small-scale active injection leads to different dynamical behaviors ranging from a quiescent state (see Fig.~\ref{fig1}a) at low energy injection rate, dominated by elastic relaxation, to a \emph{spontaneous flow} regime (Fig.~\ref{fig1}b), where long-ranged elastic instabilities, induced by activity, are able to produce and autonomously sustain laminar flows~\cite{Wu1979,sanchez2012,voituriez2005,Carenza22065, Negro2019_EPL} and give rise to many unexpected behaviors~\cite{loisy2018,Guo201722505,giomi2010,negro2019,cugliandolo2017,carenza2020_scirep}. 

Experiments on active gels have shown that activity may further lead to a chaotic state by strengthening
elastic deformations, until the threshold of production of topological defect pairs is reached~\cite{Decamp2015}. This regime, characterized by the development of vortical flows, is commonly --and misleadingly (as we will discuss in this Letter)-- addressed as \emph{active turbulence}~\cite{dombrowski2004,wensink2012,Doostmohammadi2017,dunkel2013,Duclos2020,Guillamat2017,carenza2020_physA,negro2018},
due to the  qualitative resemblance to hydrodynamic turbulence at high Reynolds numbers $Re$~\cite{alexakis2018}. 

The mechanism at the base of such chaotic behavior is yet not fully understood. In particular, the question concerning how (and \emph{if}) energy may be transferred among different length-scales by means of some non-linear interactions --as it happens in classic hydrodynamic turbulence-- still remains unanswered. 
The transition from the laminar towards the chaotic state has been widely analyzed in terms of the topological properties~\cite{thampi2014_2,giomi2015,Tan2019} and classified as belonging to the direct percolation universality class~\cite{Doostmohammadi2017}, for a system confined in a channel with rigid walls. Furthermore, 
the spectral properties of active gels have been investigated in  different experimental realizations~\cite{wensink2012,Bratanov2015,creppy2015,dombrowski2004} and studied by means of different numerical and analytical models~\cite{Linkmann2019,giomi2015,wolgemuth2008,thampi2013,Urzay2017,Alert2020} which often do not agree with each other and do not fit into the scenario of turbulence at high $Re$.
In particular, in a $2d$ isotropic turbulent fluid --where non-linear advection overcomes viscous dissipation-- a counter directional
dual cascade takes place with inverse energy cascade towards larger length-scales and direct enstrophy cascade to smaller ones~\cite{alexakis2018}. To speak about turbulence you need to be in presence of at least two ingredients: a highly chaotic spatio-temporal behaviour and an energy transfer (cascade) over a  controllable scale separation between injection and dissipative mechanisms.
Conversely, active gels flow at negligible $Re$ ($\lesssim 10^{-1}$), a regime where hydrodynamic advection is not likely playing a significant role,  precluding hydrodynamic  non-linearities to be responsible for the onset of the instability in dense active suspensions~\cite{dombrowski2004,Ishikawa2011}.
Still, one may expect energy to be transferred by means of non-linear elastic interactions, analogously to what happens in polymer solutions flowing at low $Re$~\cite{groisman2000,steinberg2019,morozov2007}.

A systematic approach to address the energy transfer mechanism in active gels was recently used by Urzay \emph{et al.}~\cite{Urzay2017} which showed the absence of hydrodynamic turbulence making use of the full nemato-hydrodynamic theory for active nematics. Later on, Alert \emph{et al.}~\cite{Alert2020} analytically and numerically studied the energy balance in Fourier space in a minimal model for uniaxial defect-free active nematics, finding an universal power-law scaling $ k^{-1}$ of the kinetic energy spectrum at small wave-numbers due to the long-ranged visco-elastic interactions, without any energy cascade.
The absence of hydrodynamic advection was also reported by the authors of this paper for the case of an active polar emulsion, even if no scale-invariance was found~\cite{carenza2020}.
Interestingly, Linkmann \emph{et al.}~\cite{Linkmann2020} have shown that the nemato-hydrodynamic equations used in the aforementioned papers can be mapped into the Eulerian S{\l}omka-Dunkel model~\cite{wensink2012,Bratanov2015,Slomka2015} where the range of scales and the rate of energy injection can be selected by tuning some model parameters. Controversially, this single-fluid model predicts energy to be transferred between length-scales by means of advective interactions~\cite{slomka2018,slomka2019,Linkmann2019}, thus leaving open the question if the chaotic regime found in 
the models of active gels  can be characterized as a turbulent cascade. The question is not semantic, as in the presence of turbulence we must be able to control the intensity and the scale-extension of the fluctuating fields, and we expect universal behaviour independent of the details of the forcing mechanisms. 

The goal of this paper is to study the energy transfer mechanism in a well established model for active nematics~\cite{carenza2019,Urzay2017,giomi2015}, 
in order to disentangle all possible contributions behind the complex multi-scale behaviour including the --so far-- elusive role of  elasticity. We start from the full nemato-hydrodynamic theory for active nematics and we perform a systematic spectral analysis to elucidate the role of reactive, elastic, kinematic and dissipative contributions to the multi-scale energy dynamics.
We confirm that kinematic advective contributions are factually negligible and dynamics to be mostly driven by the mutual balancing of active injection and viscous dissipation without any important turbulent energy transfer across scales, except for a small contribution given by  elasticity which moves energy from small scales towards larger ones, giving rise to an effective small non-linear inverse energy transfer. 
\\

\begin{figure}[bt]
\centering
\includegraphics[width=1.0\columnwidth]{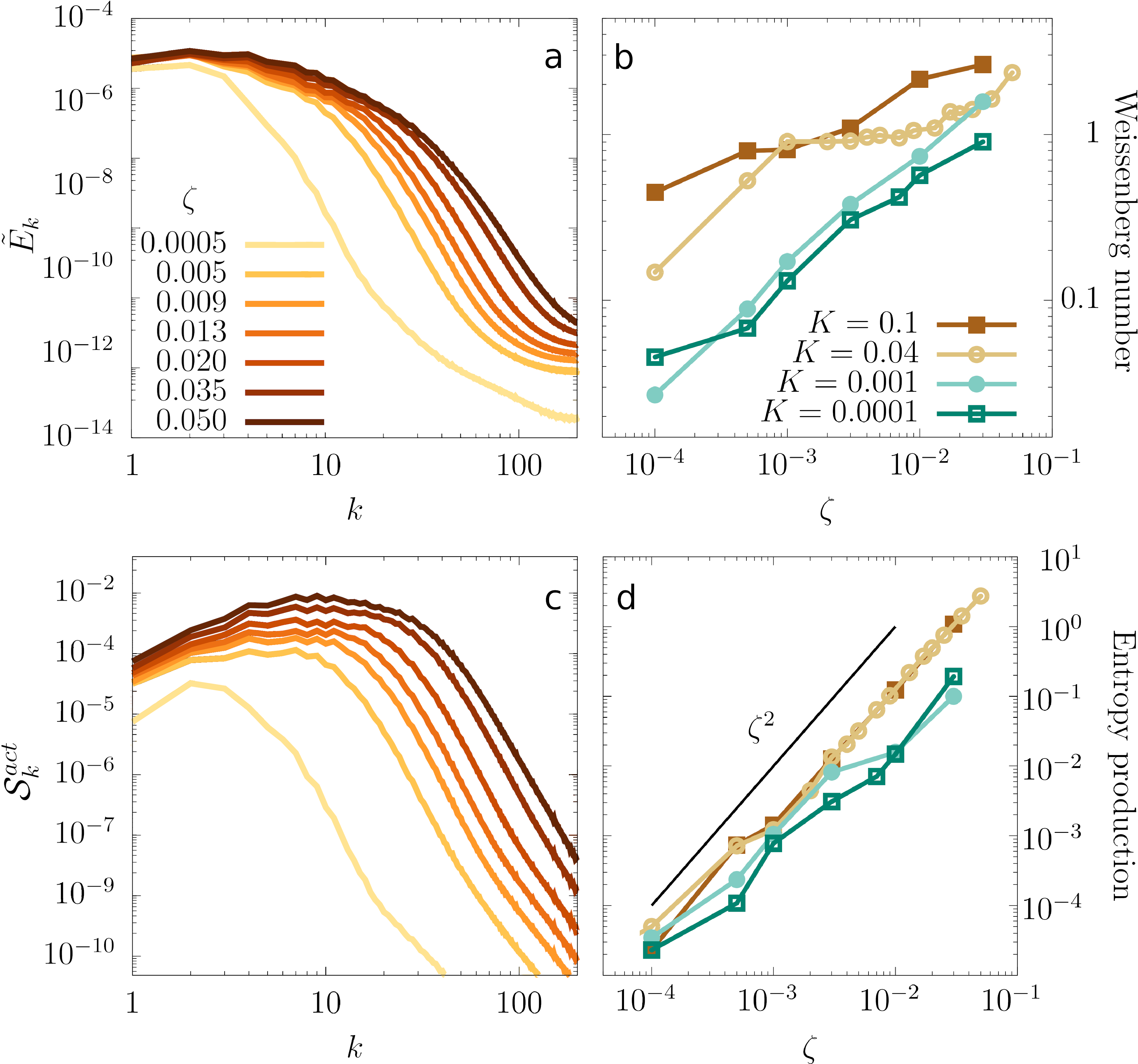}
\caption{\textbf{Spectral properties of the turbulent state.} Panel~(a) shows the normalized energy spectra $\tilde{E}_k= E_k U_O / (l_a^2 \int dk \ S^{act}_k)$ for some values of the activity parameter $\zeta$ and $K=0.04$, where $U_0$ is the typical flow velocity (see text). Panel~(c) shows the corresponding spectra of active energy input $ S^{act}_k$. Panels~(b) and~(d) show  the continuous behavior respectively of the Weissenberg number and the entropy production at varying $\zeta$ for different values of the elastic constant $K$. Simulations are performed on a square grid of size $L=512$.}
\label{fig2}
\end{figure}

\textbf{Model} 
We model the dynamics of a bidimensional active gel by means of the Landau-De Gennes theory for LC~\cite{degennes1993}. We consider $\rho$ and $\mathbf{v}$, respectively the density and the velocity of the fluid. The ordering properties of the LC are encoded in the nematic (trace-less and symmetric) tensor $Q_{\alpha \beta}$, whose principle eigenvector $\mathbf{n}$  --the director-- defines the local preferential direction of alignment of the LC. 
The dynamics of the active gel is ruled by the following set of equations:
\begin{equation}
(\partial_t + \mathbf{v}\cdot \nabla) \mathbf{Q} - \mathbf{S}(\nabla \mathbf{v},\mathbf{Q}) = \Gamma^{-1} \mathbf{H}.  \label{eqn:BSE}
\end{equation}
\begin{equation}
\rho (\partial_t + \mathbf{v} \cdot \nabla ) \mathbf{v} = -\nabla p + \nabla \cdot \left[ \mathbf{\sigma}^{pass} + \mathbf{\sigma}^{act} \right], \label{eqn:NSE}
\end{equation}
The first is the Beris-Edwards equation which defines the relaxation of the LC. Here $\Gamma=1$ is the rotational viscosity, while
\begin{multline*}
\mathbf{S}(\nabla \mathbf{v},\mathbf{Q}) = (\xi \mathbf{D} + \mathbf{\Omega})(\mathbf{Q}+\mathbf{I}/3) \\ + (\mathbf{Q}+\mathbf{I}/3)(\xi \mathbf{D}  - \mathbf{\Omega}) - 2 \xi (\mathbf{Q}+\mathbf{I}/3) Tr (\mathbf{Q} \nabla \mathbf{v}),
\end{multline*}
is the strain rotational derivatives where $D_{ \alpha \beta}= (\partial_\alpha v_\beta + \partial_\beta v_\alpha)/2$ is the strain rate tensor, $\Omega_{ \alpha \beta}= (\partial_\alpha v_\beta - \partial_\beta v_\alpha)/2$ is the vorticity tensor and $\xi$ is the dimensionless alignment parameter, related to the shape of the suspended particles. Unless otherwise stated, we choose $\xi=0.7$, corresponding to flow aligning rods~\cite{marenduzzo2007}.
The traceless molecular field $ H_{ij}= -\left(\frac{\delta \mathcal{F}}{\delta Q_{ij}}\right)^{tl}$ is derived from the Landau-De Gennes free energy $\mathcal{F} = \int d \bm{r} f$, with the free energy density $f$  given by the sum of a bulk contribution $f^{bulk}=A_0 \left[ \frac{1}{2} \left(1 - \frac{\chi}{3} \right)\mathbf{Q}^2 -  \frac{\chi}{3} \mathbf{Q}^3 +  \frac{\chi}{4} \mathbf{Q}^4 \right]$ and an elastic term $f^{el}= \frac{K}{2} (\nabla  \mathbf{Q})^2 $, where $A_0=0.8$ and $K$ are respectively the bulk and elastic constant and $\chi$ is a parameter controlling the isotropic-nematic transition, occurring when $\chi>2.7$~\cite{degennes1993}.
Eq.~\eqref{eqn:NSE} is the Navier-Stokes equation,
where $p$ is the ideal fluid pressure and the stress tensor has been divided in a \emph{passive} ($\sigma^{pass}$) and an \emph{active} ($\sigma^{act}$) part.
The former accounts for dissipative and reactive effects and can be expressed as the sum  of a viscous contribution $\sigma^{visc}= 2 \eta \mathbf{D}$, with $\eta$ the fluid viscosity, and an elastic one
\begin{multline}
\sigma_{\alpha \beta}^{el} = -\xi H_{\alpha \gamma} \left(Q_{\gamma \beta} + \dfrac{1}{3} \delta_{\gamma \beta} \right) -\xi  \left(Q_{\alpha \gamma} + \dfrac{1}{3} \delta_{\alpha \gamma} \right) H_{\gamma \beta} \\ 	 + 2\xi \left(Q_{\alpha \beta} - \dfrac{1}{3} \delta_{\alpha \beta} \right) Q_{\gamma \mu} H_{\gamma \mu} + Q_{\alpha \gamma} H_{\gamma \beta}  - H_{ \alpha \gamma} Q_{\gamma \beta} . 
\label{eqn:elastic_st}
\end{multline}
The active stress is given by $\sigma^{act}=-\zeta \mathbf{Q}$~\cite{pedley1992,hatwalne2004}. The constant $\zeta$, the activity, tunes the intensity of the active doping and describes extensile particles, if $\zeta>0$, or contractile ones otherwise.

We numerically integrate Eqs.~(\ref{eqn:BSE}) and~(\ref{eqn:NSE}) on a 2D simulation box of size $L=512$ by means of a hybrid lattice Boltzmann method~\cite{carenza2019}, so that the fluid may be in principle slightly compressible. However, the Mach number $Ma=\bar{v}/c_s \ll 1$ in our simulations (where $\bar{v}$ denotes the average fluid velocity and $c_s$ the speed of sound) and density fluctuations $\delta \rho ~\sim Ma^2$, are to all practical effect negligible. Therefore, in the following we assume density to be homogeneous ($\rho=1$) so that the flow field satisfies the condition of incompressibility $\nabla \cdot \mathbf{v}=0$.
Simulations units can be mapped onto physical ones by fixing the grid spacing $\Delta x = 5 \mu m $, the time-step $\Delta t= 20 ms $  and the force-scale $f^* =2 \mu N$. In our simulations the viscosity is set to $\eta=5/6$ corresponding to  $1.33 \textit{kPas}$.


\begin{figure}[bt]
\centering
\includegraphics[width=1.0\columnwidth]{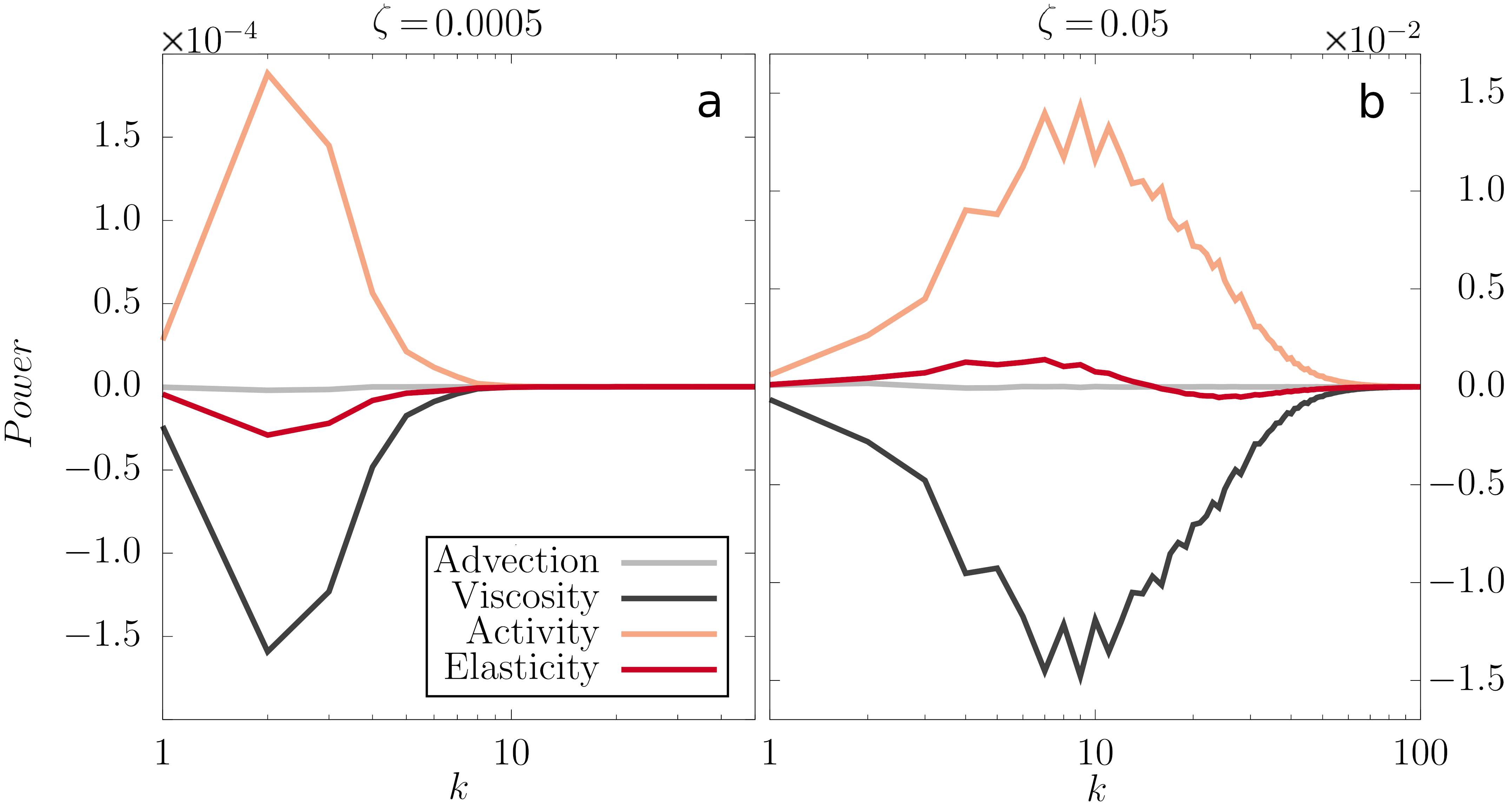}
\caption{\textbf{Energy balance in Fourier space.} Time averaged power spectra at $K=0.04$, for the spontaneous flow regime at $\zeta=5 \times 10^{-4}$ are shown in panel~(a) and for the chaotic regime at $\zeta=5 \times 10^{-2}$ in panel~(b). Notice that the strength of active injection (and viscous dissipation) increases by a factor $\sim 10^2$.}
\label{fig3}
\end{figure}

\begin{figure}[bt]
\centering
\includegraphics[width=1.0\columnwidth]{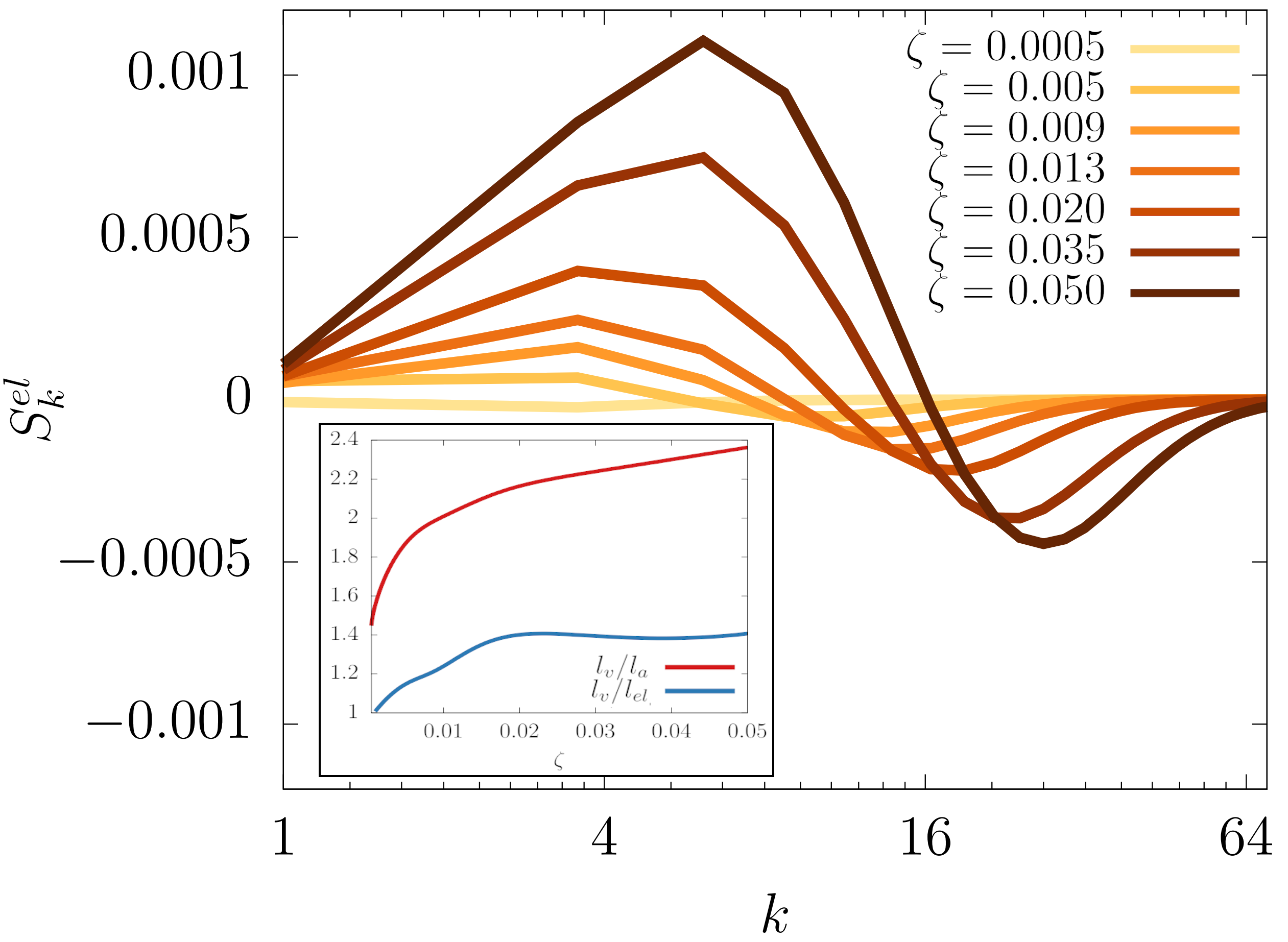}
\caption{\textbf{Elastic energy transfer.} Elastic spectra $ S^{el}_k$ for some values of $\zeta$ and $K=0.04$. The color scale is the same as in Fig.~\ref{fig2}. The red (blue) line in the inset shows the ratio between the typical flow scale $l_v$ and the active injection $l_a$ (elastic $l_{el}=L/k_{el}$) length-scale.}
\label{fig4}
\end{figure}

\section{Our results}
We vary the intensity of extensile activity and the elastic constant of the LC, moving from the quiescent regime of Fig.~\ref{fig1}(a) to the chaotic case shown in panel~(c). 
It is known that the dynamical equations~(\ref{eqn:BSE}) and~(\ref{eqn:NSE}) exhibit an elastic instability to bending deformations at increasing the activity parameter $\zeta$~\cite{simha2002, kruse2004,giomi2015}. This first occurs when the active length-scale $l_a \sim \sqrt{K/\zeta}$ drops under half the size of the system, being able to excite a long wave-length modulation of the nematic pattern, as shown in Fig.~\ref{fig1}(b) for the case at $\zeta=5 \times 10^{-4}$. The deformation of the LC injects energy in the system by means of the active stress $\sigma^{act}$ in the direction normal to the bending. 
Our approach to quantitatively study the response of elastic and dissipative contributions to active injection is to consider a balance equation for the kinetic energy in Fourier space:
\begin{equation}
\partial_t E_k + \mathcal{T}_k = \mathcal{S}^{visc}_k + \mathcal{S}^{el}_k + \mathcal{S}^{act}_k.
\label{eqn:energy}
\end{equation}
Here $E_k=\langle  | \mathbf{v}_\mathbf{k}|^2 \rangle /2$ is the energy spectrum, $\langle \cdot \rangle$ denoting the sum on shells of equal momentum ($|\mathbf{k}|=k$), and $\mathcal{T}_k=  \langle \mathbf{v}_\mathbf{k}^* \cdot \mathbf{J}_\mathbf{k} \rangle$ is the rate at which energy is transferred by non-linear advective interactions, with $\mathbf{J}_\mathbf{k}$ the Fourier representation of the hydrodynamic flux $-\nabla p + \mathbf{v}\cdot \nabla \mathbf{v}$.
The terms on the right-hand side of Eq.~(\ref{eqn:energy}) are respectively given by
$$
\mathcal{S}^{visc}_k= 2 \pi i \langle \mathbf{v}_\mathbf{k}^* \otimes \mathbf{k} : \sigma^{visc}_\mathbf{k} \rangle/L = - \frac{8 \pi^2 \eta}{L^2}  k^2 E_k
$$
$$
\mathcal{S}^{el}_k= 2 \pi i \langle \mathbf{v}_\mathbf{k}^* \otimes \mathbf{k} : \sigma^{el}_\mathbf{k} \rangle/L 
$$
$$
\mathcal{S}^{act}_k= 2 \pi i \langle \mathbf{v}_\mathbf{k}^* \otimes \mathbf{k} : \sigma^{act}_\mathbf{k} \rangle/L 
$$
and represent the rate at which energy is absorbed, injected or --eventually-- transferred among length-scales, by viscous, elastic and active contributions.

By comparing the energy spectrum $E_k$ and the spectrum of active injection $\mathcal{S}^{act}_k$, respectively shown in Fig.~\ref{fig2}(a) and~(c) for the case with smallest activity (spontaneous flow regime) at $\zeta=5 \times 10^{-4}$, we observe that the flow develops fluctuations at the same  length-scale where energy is injected, giving rise to the laminar flow observed in panel~(b) of Fig.~\ref{fig1}. 
As activity increases, the active stress injects energy at larger and larger wave-numbers (see Fig.~\ref{fig2}(c)), thus strengthening the bending  of the active nematics, leading to the presence of narrow regions, \emph{walls}~\cite{thampi2014_1,giomi2015}, characterized by strong deformations which eventually produce a  proliferation of topological defects (point-like disclinations). These are singular regions where the nematic order is lost and it is not possible to define the mean orientation of the LC molecules. Walls and disclinations play a relevant role on the onset of the chaotic regime since the strong distortions in their neighborhood generate flows which are in turn responsible for the deformation of the walls and the unbinding of more defects pairs~\cite{giomi2014_2,Shendruk2017}, thus driving the system towards the apparently \emph{turbulent} state.
The increasing amount of energy injected in the system has the important effect of strengthening the flows, which develop on a wider range of scales, as suggested by the behavior of the energy spectra at large activity in Fig.~\ref{fig2}(a). 
In order to characterize the hydrodynamic response of the active nematic to elastic deformations, we introduce an active Weissenberg number $Wi=U_0 \tau_{el}/l_a$~\cite{steinberg2019}, where $U_0=\sqrt{\int dk E_k}$ is the typical flow velocity, $\tau_{el}=  U_0^2 / (\int dk |\mathcal{S}^{el}_k|)$ is the relaxation time of the LC and  $l_a=L/k_a$ is the typical length-scale of active injection, with $k_a=(\int dk \mathcal{S}^{act}_k k)/(\int dk \mathcal{S}^{act}_k)$. 
The behavior of $Wi$ at varying $\zeta$ (see Fig.~\ref{fig2}(b)) reflects the continuous nature of the transition from the spontaneous flow towards the chaotic regime, without any singularity neither in the hydrodynamic nor in the energetic properties of the system, regardless of the strength of the LC elasticity. This is also confirmed  by the power-law scaling $\sim \zeta^2$ of the entropy production $s=\eta (\nabla \mathbf{v}^{tl})^2+ \frac{1}{\Gamma} \mathbf{H}^2$, shown in Fig.~\ref{fig2}(d). 



\begin{figure}[bt]
\centering
\includegraphics[width=1.0\columnwidth]{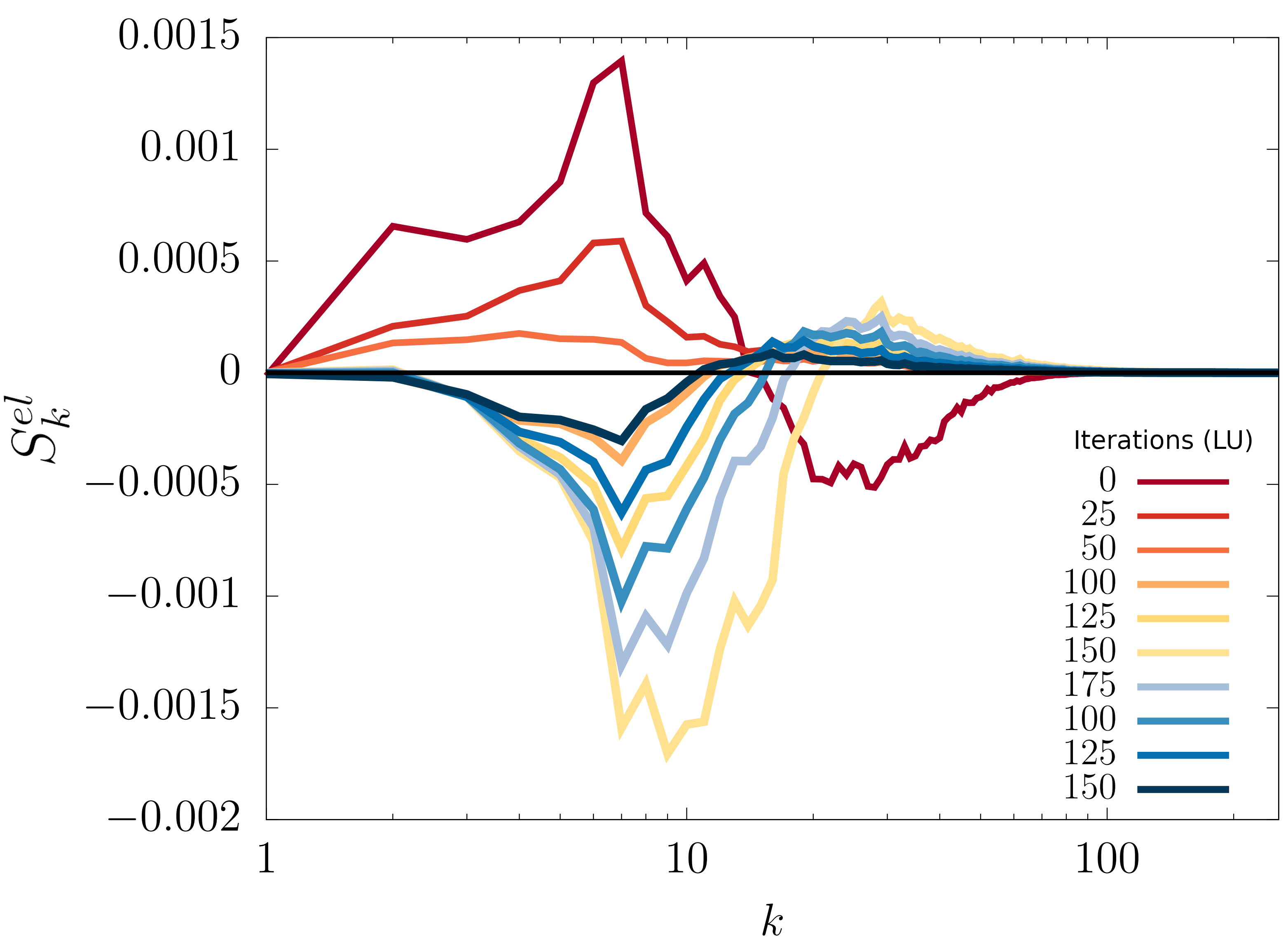}
\caption{\textbf{Elastic relaxation.} Dynamical response of the elastic spectrum $ S^{el}_k$ as activity is switched off starting from a chaotic steady-state configuration at $\zeta=5 \times 10^{-2}$. The time iterations are reported in the legend in lattice units.
}
\label{fig5}
\end{figure}

As shown by the previous discussion, energy spectra do not provide enough information to disentangle the intricate transfer mechanism in a complex fluid. In particular, it is not possible on the basis of the spectrum only to understand what are the physical mechanisms behind the formation of given structures at a given scale. In order to achieve it, we analyzed the energy balance in Fourier space by considering the scale-to-scale contributions of the terms in Eq.~(\ref{eqn:energy}), shown in Fig.~\ref{fig3}.
We observe that in the spontaneous flow regime (left panel, corresponding to Fig.~\ref{fig1}(b)) only small wave-numbers contribute to dynamics. This is because activity is only able to excite smooth long-ranged deformations of the LC pattern which set up and autonomously maintain laminar flows spanning the whole system.
The energy injected by activity (ochre line) is either dissipated by viscosity (gray) or used to sustain deformations in the LC pattern (red), leading to a localized balance, scale-by-scale, for the energy dynamics.

As $\zeta$ is increased and the system enters the chaotic regime, the rate of active injection considerably increases as activity excites modes at larger and larger wave-numbers, while the energy scale of elastic effects remains roughly unaltered, as suggested by a direct comparison of the two cases shown in Fig.~\ref{fig3}. In agreement with previous works~\cite{Urzay2017,Alert2020},
the dynamics is basically driven by the local balancing between active injection and viscous dissipation, while hydrodynamic convective interactions are practically negligible.
Interestingly enough, we observe here that 
the elastic term change sign developing  a (small) non linear energy transfer from large towards small wave-numbers (see right panel of Fig.~\ref{fig3}), due to cross-triadic interactions in $\mathcal{S}^{el}_k$ between the nematic tensor in Eq.~\eqref{eqn:elastic_st} and the velocity field, thus opening the road to have an equilibrium parameter in the system to control the flow response.
Thus, it is of interest to further analyze the behavior of the elastic contribution at varying the elastic constant $K$. For $K\lesssim 10^{-2}$, the elastic term $\mathcal{S}^{el}_k$ counters active injection, regardless of the intensity of the active pumping, by absorbing energy at any length-scale, analogously to what happens in the spontaneous flow regime. This picture drastically changes when stiffer LC ($K\gtrsim 10^{-2}$) are considered. 
In this case, the behavior of the elastic contribution develops a positive branch at small wave-numbers and a negative one at large $k$  (Fig.~\ref{fig3}(b) and Fig.~\ref{fig4}), giving rise to an effective energy transfer from small towards larger length-scales. Fig.~\ref{fig4} shows the elastic spectrum $\mathcal{S}^{el}_k$ for $K=0.04$ for some values of $\zeta$. We observe that the non-linear energy transfer sets up as $\zeta \gtrsim 5 \times 10^{-2}$, corresponding to the threshold of the chaotic regime for the specific value of the elastic constant here considered. The amount of energy transferred among scales grows with $\zeta$ and the elastic term behaves as an energy source, since the positive branch at small wave-numbers is always larger than its negative counterpart at larger $k$.
Interestingly the wave-number $k_{el}$ defining the crossover between the source and the absorbent branch of the elastic spectrum grows towards larger $k$ as activity is increased together with the injection length-scale $l_a$ (see Fig.~\ref{fig4}). 
Moreover, we observe that the ratio between the typical flow scale $l_v=L (\int dk E_k/(\int dk E_k k)$ and the injection scale slowly increases with $\zeta$, contrary to what happens in fully developed $2d$ hydrodynamic turbulence, where the separation between the two length-scales can grow indefinitely~\cite{alexakis2018}. 
Hence, the elastic non-linear term moves an  amount of energy which is \emph{small}, if compared with the other contribution, over a limited range of scales so that energy is \emph{mostly} dissipated at the same scale where it is injected. Moreover, if the elastic transfer would play a relevant role, one would expect the statistical properties of the flow to develop scale-invariant features --\emph{i.e.} a power low decay of the energy spectrum as in hydrodynamic turbulence-- in contrast with our observations.
These results suggest that the elastic term does not establish an energy cascade even contributing to the overall dynamics with a small non-linear transfer.
This is the result --more than the cause-- of the tendency of the LC to relax deformations induced by activity at small scales ($\sim l_a < l_{el}=L/k_{el}$), giving rise to the typical configuration of Fig.~\ref{fig1}(c) where narrow \emph{walls} of strong bend deformations in the nematic pattern are spaced by wider aligned regions.


The role of the elasticity is confirmed  by a dynamical experiment consisting in observing the relaxing dynamics of the LC. We start from the fully chaotic configuration of Fig.~\ref{fig1}(c) at $\zeta=5 \times 10^{-2}$ and we switch the activity off to track the evolution of the elastic spectrum $\mathcal{S}^{el}_k$ during relaxation, as shown in Fig.~\ref{fig5}.
As activity does not support any energy input anymore, the negative branch at large wave-numbers immediately fades away. The only energy source is the elastic deformations of the nematic pattern, which progressively relax into an aligned equilibrium configuration. In this case, we observe an inversion of the behavior of the elastic spectrum with a negative branch at small $k$ which removes energy from large-scales to transfer it towards smaller scales, where it is dissipated by viscosity.

\section{Conclusion}
In this Letter we have numerically analyzed the continuous transition from spontaneous flow towards the chaotic regime of an active nematics. We found that the  transition is driven by activity by exciting distortions at larger and larger wave-numbers.
By means of a scale-to-scale  analysis, we showed that non-linear hydrodynamic interactions do not influence the dynamics of the active gel which is basically ruled by the local balancing between active energy injection and viscous dissipation, meaning that energy is mostly dissipated at those length-scales at which it has been injected, in line with the small Reynolds number $Re$, measured from simulation which never exceeds $0.1$.
This is in strict contrast with the results obtained by means of a S{\l}omka-Dunkel approach~\cite{Bratanov2015,Linkmann2019,Linkmann2020}, where the turbulent regime has hydrodynamic origin as it arises from an inverse turbulent cascade triggered by an \emph{ad hoc} linear instability in the NSE which leads to small scale injection through the action of a hyper-viscous term. However, in this case the Reynolds number $Re$ is consistently greater than $1$--a regime where inertial effects actually overcome viscous dissipation.
The energy spectra in our simulations do not exhibit power-law scaling, contrary to what was reported by Alert \emph{et al.} in~\cite{Alert2020} for the simplified case of an uniaxial defect-free active nematics. This may be due to the fact that biaxial fluctuations in the ordering properties of the LC may consistently alter the long-ranged interactions in the chaotic state, leading to the loss of scale invariance --an aspect which deserves further investigation. 
Furthermore, we found that the non-linear elastic terms may qualitatively play an important role by moving energy between different scales, developing a positive branch at small wave-numbers and a negative one at large $k$, thus leading to non-linear energy transfer. The rate at which energy is transferred among scales, though, is small compared with the injection/dissipation rate and the elastic contributions cannot sustain any energy cascade.
Hence, our results show that the full nemato-hydrodynamic theory for active gels does not exhibit typical  \emph{turbulent} dynamics. Accumulation of energy at large scales occurs as the effect of non-trivial couplings between the banded patterns of the LC --which lead to active energy input-- and the velocity field. 
Nonetheless, the question whether it exists a range of parameters where it is actually possible to observe a \emph{fully developped} inverse cascade still remains open. In such a case we expect the elastic energy transfer to drive the dynamics of the system and eventually develop power-law scaling in an extended range of scales comprised between the small scales of active injection and the larger scales where the velocity field develops vortical structures.


\bibliographystyle{unsrt}
\bibliography{biblio}

\begin{thebibliography}{10}

\bibitem{ramaswamy2010}
S.~Ramaswamy.
\newblock The mechanics and statistics of active matter.
\newblock {\em Annu. Rev. Condens. Matter Phys.}, 1:323, 2010.

\bibitem{marc2013}
M.C. Marchetti, J.F. Joanny, S.~Ramaswamy, T.B. Liverpool, J.~Prost, M.~Rao,
  and R.A. Simha.
\newblock Hydrodynamics of soft active matter.
\newblock {\em Rev. Mod. Phys}, 85:1143, 2013.

\bibitem{wensink2012}
H.H. Wensink, J.~Dunkel, S.~Heidenreich, K.~Drescher, H.~Lowen R.E.~Goldstein,
  and J.M. Yeomans.
\newblock Meso-scale turbulence in living fluids.
\newblock {\em Proc. Natl. Acad. Sci.}, 109, 2012.

\bibitem{guillamat2018}
P.~Guillamat, Z.~Kos, J.~Hardo\"uin, M.~Ravnik, and R.~Sagu\'es.
\newblock Active nematic emulsions.
\newblock {\em Science Advances}, 4:4, 2018.

\bibitem{ebbens2014}
S.~Ebbens, D.A. Gregory, G.~Dunderdale, J.R. Howse, Y.~Ibrahim, T.B. Liverpool,
  and R.~Golestanian.
\newblock Electrokinetic effects in catalytic platinum-insulator janus
  swimmers.
\newblock {\em Europhys. Lett.}, 106:5, 2014.

\bibitem{korevaar2020}
P.A. Korevaar, C.N. Kaplan, A.~Grinthal, R.M. Rust, and J.~Aizenberg.
\newblock Non-equilibrium signal integration in hydrogels.
\newblock {\em Nat. Comm.}, 11:386, 2010.

\bibitem{sanchez2012}
T.~Sanchez, D.T.N. Chen, S.J. Decamp, M.~Heymann, and Z.~Dogic.
\newblock Spontaneous motion in hierarchically assembled active matter.
\newblock {\em Nature}, 491:431--434, 2012.

\bibitem{kruse2004}
K.~Kruse, J.F. Joanny, F.~J\"ulicher, J.~Prost, and K.~Sekimoto.
\newblock Asters, vortices, and rotating spirals in active gels of polar
  filaments.
\newblock {\em Phys. Rev. Lett.}, 92:078101, 2004.

\bibitem{Wu1979}
K.-T. Wu, J.B. Hishamunda, D.T.N. Chen, S.J.~De Camp, Y.-W. Chang,
  A.~Fern{\'a}ndez-Nieves, S.~Fraden, and Z.~Dogic.
\newblock Transition from turbulent to coherent flows in confined
  three-dimensional active fluids.
\newblock {\em Science}, 355(6331), 2017.

\bibitem{voituriez2005}
R.~Voituriez, J.F. Joanny, and J.~Prost.
\newblock Spontaneous flow transition in active polar gels.
\newblock {\em E.P.L.}, 70:404, 2005.

\bibitem{Carenza22065}
L.N. Carenza, G.~Gonnella, D.~Marenduzzo, and G.~Negro.
\newblock Rotation and propulsion in 3d active chiral droplets.
\newblock {\em Proc. Natl. Acad. Sci.}, 116(44):22065--22070, 2019.

\bibitem{Negro2019_EPL}
G.~Negro, A.~Lamura, G.~Gonnella, and D.~Marenduzzo.
\newblock Hydrodynamics of contraction-based motility in a compressible active
  fluid.
\newblock {\em EPL}, 127(5):58001, 2019.

\bibitem{loisy2018}
A.~Loisy, J.~Eggers, and T.B. Liverpool.
\newblock Active suspensions have nonmonotonic flow curves and multiple
  mechanical equilibria.
\newblock {\em Phys. Rev. Lett.}, 121, 2018.

\bibitem{Guo201722505}
Shuo Guo, Devranjan Samanta, Yi~Peng, Xinliang Xu, and Xiang Cheng.
\newblock Symmetric shear banding and swarming vortices in bacterial
  superfluids.
\newblock {\em Proc. Natl. Acad. Sci.}, 2018.

\bibitem{giomi2010}
L.~Giomi, T.~B. Liverpool, and M.~C. Marchetti.
\newblock Sheared active fluids: Thickening, thinning, and vanishing viscosity.
\newblock {\em Phys. Rev. E}, 81:051908, 2010.

\bibitem{negro2019}
G.~Negro, L.N. Carenza, A.~Lamura, A.~Tiribocchi, and G.~Gonnella.
\newblock Rheology of active polar emulsions: from linear to unidirectional and
  unviscid flow, and intermittent viscosity.
\newblock {\em Soft Matter}, 15:8251--8265, 2019.

\bibitem{cugliandolo2017}
L.F. Cugliandolo, P.~Digregorio, G.~Gonnella, and A.~Suma.
\newblock Phase coexistence in two-dimensional passive and active dumbbell
  systems.
\newblock {\em Phys. Rev. Lett.}, 119:268002, 2017.

\bibitem{carenza2020_scirep}
L.N. Carenza, G.~Gonnella, A.~Lamura, D.~Marenduzzo, G.~Negro, and
  T.~Tiribocchi.
\newblock Soft channel formation and symmetry breaking in exotic active
  emulsions.
\newblock {\em Sci. Rep.}, (on press), 2020.

\bibitem{Decamp2015}
S.J. DeCamp, G.S. Redner, A.~Baskaran, M.F. Hagan, and Z.~Dogic.
\newblock Orientational order of motile defects in active nematics.
\newblock {\em Nat. Mater.}, 14:11110, 2015.

\bibitem{dombrowski2004}
C.~Dombrowski, L.~Cisneros, S.~Chatkaew, R.E. Goldstein, and J.O. Kessler.
\newblock Self-concentration and large-scale coherence in bacterial dynamics.
\newblock {\em Phys. Rev. Lett.}, 93:098103, 2004.

\bibitem{Doostmohammadi2017}
A.~Doostmohammadi, T.N. Shendruk, K.~Thijssen, and J.M. Yeomans.
\newblock Onset of meso-scale turbulence in active nematics.
\newblock {\em Nat. Comm.}, 8, 2017.

\bibitem{dunkel2013}
J.~Dunkel, S.~Heidenreich, K.~Drescher, H.H. Wensink, M.~B\"ar, and R.E.
  Goldstein.
\newblock Fluid dynamics of bacterial turbulence.
\newblock {\em Phys. Rev. Lett.}, 110:228102, 2013.

\bibitem{Duclos2020}
G.~Duclos, R.~Adkins, D.~Banerjee, M.S.E. Peterson, M.~Varghese, I.~Kolvin,
  A.~Baskaran, R.A. Pelcovits, T.R. Powers, A.~Baskaran, F.~Toschi, M.F. Hagan,
  S.J. Streichan, V.~Vitelli, D.A. Beller, and Z.~Dogic.
\newblock Topological structure and dynamics of three-dimensional active
  nematics.
\newblock {\em Science}, 367(6482):1120--1124, 2020.

\bibitem{Guillamat2017}
P.~Guillamat, J.~Ign{\'e}s-Mullol, and F.~Sagu{\'e}s.
\newblock Taming active turbulence with patterned soft interfaces.
\newblock {\em Nat. Comm.}, 8, 12 2017.

\bibitem{carenza2020_physA}
L.N. Carenza, G.~Gonnella, D.~Marenduzzo, and G.~Negro.
\newblock Chaotic and periodical dynamics of active chiral droplets.
\newblock {\em Physica A}, 559:125025, 2020.

\bibitem{negro2018}
G.~Negro, L.N. Carenza, P.~Digregorio, G.~Gonnella, and A.~Lamura.
\newblock Morphology and flow patterns in highly asymmetric active emulsions.
\newblock {\em Physica A: Statistical Mechanics and its Applications}, 503:464
  -- 475, 2018.

\bibitem{alexakis2018}
A.~Alexakis and L.~Biferale.
\newblock Cascades and transitions in turbulent flows.
\newblock {\em Phys. Rep.}, 767-769:1 -- 101, 2018.
\newblock Cascades and transitions in turbulent flows.

\bibitem{giomi2014_2}
L.~Giomi, M.J. Bowick, P.~Mishra, R.~Sknepnek, and M.C. Marchetti.
\newblock Defect dynamics in active nematics.
\newblock {\em Philos. Trans. Royal Soc. A}, 372, 2014.

\bibitem{thampi2014_1}
S.P. Thampi, R.~Golestanian, and J.M. Yeomans.
\newblock Instabilities and topological defects in active nematics.
\newblock {\em EPL}, 105(1):18001, 2014.

\bibitem{giomi2011}
L.~Giomi, L.~Mahadevan, B.~Chakraborty, and M.F. Hagan.
\newblock Excitable patterns in active nematics.
\newblock {\em Phys. Rev. Lett.}, 106:218101, May 2011.

\bibitem{Urzay2017}
J.~Urzay, A.~Doostmohammadi, and J.M. Yeomans.
\newblock Multi-scale statistics of turbulence motorized by active matter.
\newblock {\em Journal of Fluid Mechanics}, 822, 07 2017.

\bibitem{bonelli2019}
F.~Bonelli, L.N. Carenza, G.~Gonnella, D.~Marenduzzo, E.~Orlandini, and
  A.~Tiribocchi.
\newblock Lamellar ordering, droplet formation and phase inversion in exotic
  active emulsions.
\newblock {\em Sci. Rep.}, 9:2801, 2019.

\bibitem{thampi2014_2}
S.P. Thampi, R.~Golestanian, and J.M. Yeomans.
\newblock Vorticity, defects and correlations in active turbulence.
\newblock {\em Philosophical Transactions of the Royal Society of London A:
  Mathematical, Physical and Engineering Sciences}, 372(2029), 2014.

\bibitem{giomi2015}
L.~Giomi.
\newblock Geometry and topology of turbulence in active nematics.
\newblock {\em Physical Review X}, 5, 2015.

\bibitem{Tan2019}
A.J. Tan, E.~Roberts, S.A. Smith, U.A. Olvera, J.~Arteaga, S.~Fortini, K.A.
  Mitchell, and L.S. Hirst.
\newblock Topological chaos in active nematics.
\newblock {\em Nat. Phys.}, 8 2019.

\bibitem{Bratanov2015}
V.~Bratanov, F.~Jenko, and E.~Frey.
\newblock New class of turbulence in active fluids.
\newblock {\em Proc. Natl. Acad. Sci.}, 112(49):15048--15053, 2015.

\bibitem{creppy2015}
A.~Creppy, O.~Praud, X.~Druart, P.L. Kohnke, and F.~Plourabou\'e.
\newblock Turbulence of swarming sperm.
\newblock {\em Phys. Rev. E}, 92:032722, 2015.

\bibitem{Linkmann2019}
M.~Linkmann, G.~Boffetta, M.C. Marchetti, and B.~Eckhardt.
\newblock {Phase Transition to Large Scale Coherent Structures in
  Two-Dimensional Active Matter Turbulence}.
\newblock {\em Phys. Rev. Lett.}, 122:214503, 2019.

\bibitem{wolgemuth2008}
C.W. Wolgemuth.
\newblock Collective swimming and the dynamics of bacterial turbulence.
\newblock {\em Biophys. J.}, 95(4):1564 -- 1574, 2008.

\bibitem{thampi2013}
S.P. Thampi, R.~Golestanian, and J.M. Yeomans.
\newblock Velocity correlations in an active nematic.
\newblock {\em Phys. Rev. Lett.}, 111:118101, Sep 2013.

\bibitem{Alert2020}
R.~Alert, J.-F. Joanny, and J.~Casademunt.
\newblock Universal scaling of active nematic turbulence.
\newblock {\em Nat. Phys.}, 3 2020.

\bibitem{Ishikawa2011}
T.~Ishikawa, N.~Yoshida, H.~Ueno, M.~Wiedeman, Y.~Imai, and T.~Yamaguchi.
\newblock Energy transport in a concentrated suspension of bacteria.
\newblock {\em Phys. Rev. Lett.}, 107:028102, Jul 2011.

\bibitem{groisman2000}
A.~Groisman and V.~Steinberg.
\newblock Elastic turbulence in a polymer solution flow.
\newblock {\em Nature}, 405:53--55, 2000.

\bibitem{steinberg2019}
V.~Steinberg.
\newblock Scaling relations in elastic turbulence.
\newblock {\em Phys. Rev. Lett.}, 123:234501, 2019.

\bibitem{morozov2007}
A.N. Morozov and W.~van Saarloos.
\newblock An introductory essay on subcritical instabilities and the transition
  to turbulence in visco-elastic parallel shear flows.
\newblock {\em Phys. Rep.}, 447(3):112 -- 143, 2007.

\bibitem{carenza2020}
L.N. Carenza, L.~Biferale, and G.~Gonnella.
\newblock Multiscale control of active emulsion dynamics.
\newblock {\em Phys. Rev. Fluids}, 5:011302, Jan 2020.

\bibitem{Linkmann2020}
M.~Linkmann, M.C. Marchetti, G.~Boffetta, and B.~Eckhardt.
\newblock Condensate formation and multiscale dynamics in two-dimensional
  active suspensions.
\newblock {\em Phys. Rev. E}, 101:022609, Feb 2020.

\bibitem{Slomka2015}
J.~S{\l}omka and J.~Dunkel.
\newblock Generalized navier-stokes equations for active suspensions.
\newblock {\em Eur. Phys. J.}, 224, 7 2015.

\bibitem{slomka2018}
J.~S{\l}omka, P.~Suwara, and J.~Dunkel.
\newblock The nature of triad interactions in active turbulence.
\newblock {\em J. Fluid Mech.}, 841:702–731, 2018.

\bibitem{slomka2019}
J.~S{\l}omka and J.~Dunkel.
\newblock Spontaneous mirror-symmetry breaking induces inverse energy cascade
  in 3d active fluids.
\newblock {\em Proc. Natl. Acad. Sci.}, 114(9):2119--2124, 2017.

\bibitem{carenza2019}
L.N. Carenza, G.~Gonnella, A.~Lamura, G.~Negro, and A.~Tiribocchi.
\newblock Lattice boltzmann methods and active fluids.
\newblock {\em Eur. Phys. J. E}, 42(6):81, 2019.

\bibitem{degennes1993}
P.G. de~Gennes and J.~Prost.
\newblock {\em The physics of liquid crystals}.
\newblock The International series of monographs on physics 83 Oxford science
  publications. Oxford University Press, 2nd ed edition, 1993.

\bibitem{marenduzzo2007}
D.~Marenduzzo, E.~Orlandini, M.E. Cates, and J.M. Yeomans.
\newblock Steady-state hydrodynamic instabilities of active liquid crystals:
  Hybrid lattice boltzmann simulations.
\newblock {\em Phys. Rev. E}, 76:031921, 2007.

\bibitem{pedley1992}
T.J. Pedley and J.O. Kessler.
\newblock {Hydrodynamic Phenomena in Suspensions of Swimming Microorganisms}.
\newblock {\em Annu. Rev. Fluid Mech.}, 24(1):313, 1992.

\bibitem{hatwalne2004}
Y.~Hatwalne, S.~Ramaswamy, M.~Rao, and R.A. Simha.
\newblock Rheology of active-particle suspensions.
\newblock {\em Phys. Rev. Lett.}, 92:118101, 2004.

\bibitem{simha2002}
R.A. Simha and S.~Ramaswamy.
\newblock Hydrodynamic fluctuations and instabilities in ordered suspensions of
  self-propelled particles.
\newblock {\em Phys. Rev. Lett.}, 89:058101, 2002.

\bibitem{Shendruk2017}
T.N. Shendruk, A.~Doostmohammadi, K.~Thijssen, and J.M. Yeomans.
\newblock Dancing disclinations in confined active nematics.
\newblock {\em Soft Matter}, 2017.

\end{thebibliography}

\end{document}